\documentclass[amssymb,aps,amsmath,floatfix,prl,twocolumn,superscriptaddress,longbibliography]{revtex4-1}
\usepackage[latin1]{inputenc}
\usepackage{mathptmx}
\usepackage[T1]{fontenc}
\usepackage{graphicx}
\usepackage{amsmath,amssymb,amsfonts, MnSymbol}
\usepackage{mathrsfs}
\usepackage{bm}
\usepackage{hyperref}
\usepackage{bbold}
\usepackage{xcolor}
\usepackage{tikz}
\usepackage{overpic}
\newcommand{\algn}[1]{\begin{align} #1 \end{align}}
\newcommand{\sbeqs}[1]{\begin{subequations} #1 \end{subequations}}

\newcommand{\ve}[1]{\boldsymbol{#1}}

\newcommand{\dd}[1]{\ensuremath{\tfrac{\text{d}}{\text{d} #1}}}

\newcommand{\mc}[1]{\ensuremath{\mathcal{#1}}}
\newcommand{\ms}[1]{\ensuremath{\mathscr{#1}}}
\newcommand{\mbb}[1]{\ensuremath{\mathbb{#1}}}
\newcommand{\kb}{\ensuremath{k_\text{B}}}
\newcommand{\eqnlab}[1]{\label{eq:#1}}

\newcommand{\figlab}[1]{\label{fig:#1}}
\newcommand{\eqnref}[1]{\eqref{eq:#1}}
\newcommand{\Eqnref}[1]{Eq.~\eqref{eq:#1}}
\newcommand{\Eqsref}[1]{Eqs.~\eqref{eq:#1}}

\newcommand{\figref}[1]{\ref{fig:#1}}
\newcommand{\Figref}[1]{Fig.~\ref{fig:#1}}
\newcommand{\Figsref}[1]{Figs.~\ref{fig:#1}}

\begin{document}
\title{Minimum-dissipation principle for synchronised stochastic oscillators far from equilibrium}
\author{Jan Meibohm}
\affiliation{Technische Universit\"at Berlin, Stra\ss{}e des 17. Juni 135, 10623 Berlin, Germany}
\affiliation{Department of Mathematics, King's College London, London WC2R 2LS, United Kingdom}
\author{Massimiliano Esposito}
\affiliation{Complex Systems and Statistical Mechanics, Department of Physics and Materials Science, University of Luxembourg, L-1511 Luxembourg, Luxembourg}
\begin{abstract}
We prove a linear stability-dissipation relation (SDR) for $q$-state Potts models driven far from equilibrium by a nonconservative force. At a critical coupling strength, these models exhibit a synchronisation transition from a decoherent into a synchronised state. In the vicinity of this transition, the SDR connects the entropy production rate per oscillator to the phase-space contraction rate, a measure of stability, in a simple way. For large but finite systems, we argue that the SDR implies a minimum-dissipation principle for driven Potts models as the dynamics selects stable non-equilibrium states with least dissipation. This principle holds arbitrarily far from equilibrium, for any stochastic dynamics, and for all $q$.
\end{abstract}
\maketitle
\section{Introduction}
Both equilibrium and non-equilibrium systems may exhibit order, characterised by spontaneously broken symmetries, long-range correlations, and macroscopic structure. At equilibrium, order is strictly constrained by the laws of thermodynamics, which force it to be stable and static. By contrast, far from equilibrium, where fluctuations prevail, irreversible dynamic processes lead to the appearance of so-called dissipative structures~\cite{Pri71,Kon14}, that may occur in a much richer variety of both stationary and dynamic patterns.

The theory of dissipative structures is closely bound to irreversibility~\cite{Pri71,Kon14} and to the dissipation of entropy~\cite{Pri55,Sch76,Mou86}. For near-equilibrium steady states~\cite{For22,For23}, Prigogine's theorem~\cite{Pri55} asserts that entropy production is minimal and constant~\cite{Jiu84}. However, no such general statements hold far from equilibrium, where counter examples are known~\cite{Nic70,Lan75,Jay80}. Generalisations of minimum-dissipation principles \`a la Prigogine to far-from-equilibrium systems have been attempted by many, but are hindered by the phenomenological nature of conventional non-equilibrium thermodynamics. Despite these setbacks, non-equilibrium extremum principles in the spirit of Gibbs' maximum entropy principle have fascinated statistical physicists for almost two centuries~\cite{Kir48}.

Stochastic Thermodynamics~\cite{Sei12,Bro15,Pel21} governs mesoscopic scales where fluctuations are abundant. Within Stochastic Thermodynamics, interactions between the system and the environment are treated as random and are equipped with a consistent thermodynamic interpretation. Thermodynamic consistency then holds when a local detailed balance (LDB) condition is fulfilled~\cite{Sei12,Bro15,Pel21}. Recent developments~\cite{Fal23} allow the analysis of thermodynamically consistent, mesoscopic models in the thermodynamic limit, connecting Stochastic Thermodynamics with the classical theory of non-equilibrium thermodynamics. This connection provides previously phenomenological macroscopic laws of non-equilibrium thermodynamics with a consistent justification.

\begin{figure}
	\includegraphics[width=\linewidth]{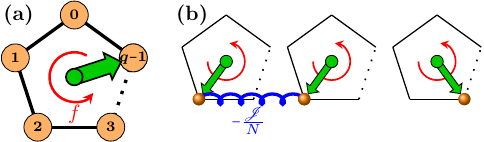}
	\caption{(a) Potts spin (green arrow) with $q$ states (orange bullets), driven by a non-conservative force $f$ (red arrow). (b) The energy of interacting Potts spins is reduced by $\ms{J}/N$ whenever two spins are aligned.}\figlab{setup}
\end{figure}
In this Letter, we employ this connection to shed new light on the old question of the existence of minimum-dissipation principles far from equilibrium. We analyse a family of thermodynamically consistent oscillators, so-called driven $q$-state Potts models~\cite{Her18,Her19}, that consist of $N$ globally interacting Potts spins. A Potts spin can be thought of as a two-dimensional unit vector, that points into one out of $q$ equally-spaced directions, see Figure~\figref{setup}(a). We denote by $s_{m}$, $m=1,\ldots,N$, the state of the $m^\text{th}$ Potts spin, which may take any integer value between $0$ and $q-1$. 

The states $s_m$ change stochastically through a dynamics that models the interactions of the system with a heat bath at inverse temperature $\beta=1/(\kb T)$. Transitions are allowed only between adjacent states $s_{m}\to s_{m}\pm1$, with $s_m$ understood modulo $q$.

The spins are driven out of equilibrium by a non-conservative force $f>0$, modelled by biasing the dynamics such that it favours counter-clockwise transitions $s_{m}\to s_{m}+1$ over clockwise transitions $s_{m}\to s_{m}-1$ [red arrow in \Figref{setup}(a)]. This way, individual Potts spins rotate on average in direction of the driving and become ``Potts oscillators''. Potential experimental realisations of driven Potts models are described in the companion paper~\cite{Mei24b}.

The oscillators interact with each other via a global, ferromagnetic potential, that reduces the energy of the system by $\ms{J}/N$ whenever two spins are in the same state, see \Figref{setup}(b).

The stochastic dynamics of the model is determined by transition rates for the transitions $s_{m}\to s_{m}\pm1$. These rates are constrained by thermodynamic consistency and symmetries, but arbitrary to some degree, which leaves one with free parameters that the model's behaviour depends upon.

The state of a large system of Potts oscillators is described by the occupation probability $\ve p(t) = (p_0,\ldots,p_{q-1})^{\sf T}$, $p_n(t)$ denoting the probability of an arbitrary Potts spin $s_m$ to be in state $s_m=n$ at time $t$. For small $\beta\!\!\ms{J}$, Potts oscillators rotate decoherently so that $\ve p(t)$ is constant and uniform, $\ve p(t) = \ve p^* \equiv (\frac1q,\ldots,\frac1q)^{\sf T}$. At a critical value $\beta\!\!\ms{J}_c$, however, a dynamical phase transition into a synchronised state occurs, in which macroscopic numbers of Potts spins oscillate in synchrony, as observed numerically for $q\leq7$ in Refs.~\cite{Her18,Her19}. The synchronised state is a simple example of a dissipative structure that spontaneously breaks time-translation symmetry and exhibits macroscopic order.

\begin{figure}
	\includegraphics[width=\linewidth]{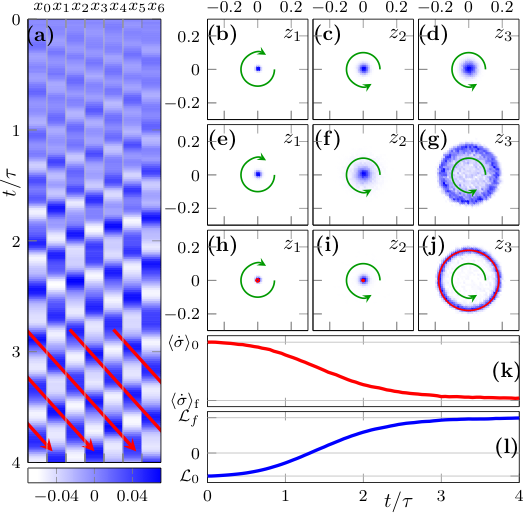}
	\caption{(a) Synchronisation in real and Fourier space starting from decoherence, using Arrhenius dynamics (details in Ref.~\cite{Mei24b}) with $q=7$, $\beta\!\!\ms{J}/q = 1.025$, $\beta f = 7$, $N=10^5$. Arrows indicate evolution in time $t$ in units of the microscopic transition time $\tau$. (a) Deviations $x_0,\ldots,x_6$ from decoherence. (b)--(j) Probability density of Fourier modes $z_1,z_2,z_3$ at $t=0.5\tau$ [(b)--(d)], $t=2\tau$ [(e)--(g)], and $t=4\tau$ [(h)--(j)], obtained from an ensemble of $10^3$ realisations. Red lines from macroscopic dynamics in \Eqnref{mf}. (k) Rate of dissipated work. (l) Phase-space contraction rate.}\figlab{oscillations}
\end{figure}

Synchronisation is conveniently described by the time-dependent deviations $\ve x(t) \equiv \ve p(t) -\ve p^*$ from decoherence. Figure~\figref{oscillations} shows synchronisation in a numerical simulation of a large system of driven Potts oscillators with $q=7$. Figure~\figref{oscillations}(a) shows how initial decoherence [$\ve x(0) = 0$] evolves into a synchronisation pattern in $\ve x(t)$, characterised by three travelling maxima (red arrows).

Discrete rotational symmetry, $s_m = s_m\pm q$, allows to characterise these patterns by discrete complex Fourier modes. As we explain below, there are three dynamical Fourier modes for $q=7$, that we denote by $z_1,z_2,z_3$ with wave numbers $k=1,2,3$, respectively. The time evolution of these modes in the complex plane is shown in \Figsref{oscillations}(b)-(j). Initially all modes are inactive [\Figsref{oscillations}(b)--(d)], but the amplitudes of the Fourier modes with largest $k$ grow [\Figsref{oscillations}(e)--(g)]. Eventually [\Figsref{oscillations}(h)--(j)], only $z_3$ is active and rotates counter-clockwise in the complex plane (green arrow), reflecting the emergence of the synchronisation pattern shown in \Figref{oscillations}(a). By contrast, the amplitudes of $z_1$ and $z_2$ are essentially zero. 

We employ Stochastic Thermodynamics~\cite{Sei12,Bro15,Pel21} to extract faithful thermodynamic observables from the dynamics. The average rate of dissipated work, shown in \Figref{oscillations}(k), is initially large and equal to $\langle \dot \sigma \rangle_0$, the average dissipation rate of a single, uncoupled oscillator. During relaxation into the synchronised state, the rate of dissipated work decreases and settles at a smaller value, $\langle\dot \sigma\rangle_\text{f}$, associated with the average dissipation rate per oscillator in the synchronised phase.

The phase-space contraction rate, shown in \Figref{oscillations}(l), is a measure of a state's momentary stability. Its initial value $\mathcal{L}_0$ is negative, reflecting that decoherent oscillations are unstable, but it increases as function of time, changes sign, and saturates at a positive value $\mathcal{L}_\text{f}>0$, indicating a stable synchronised final state. Comparing \Figsref{oscillations}(k) and (l) we observe an apparent connection between stability and dissipation.

We explore this connection and derive a stability-dissipation relation, \Eqnref{sdr}, that links least dissipation and largest stability in a simple way. Our exact derivation is based on the analytic solution of driven Potts models close to the synchronisation transition, discussed in Ref.~\cite{Mei24b}. Previous relations between stability and dissipation either refer to near-equilibrium situations~\cite{Pri71,Kon14} or lack a coherent thermodynamic interpretation~\cite{Rue96,Dae99,Sea00,Eva07}. The new relation~\eqnref{sdr}, by contrast, holds far from equilibrium and carries a transparent interpretation, as it rests upon the well-established concepts of Stochastic Thermodynamics~\cite{Sei12,Bro15,Pel21}.

On the basis of the dissipation-stability relation, we establish a minimum-dissipation principle~\cite{Pri55} for driven Potts models at large but finite $N$, that holds arbitrarily far from equilibrium, for all dynamics, and for all $q$. Existing studies of the thermodynamics of synchronisation~\cite{Imp15,Sas15,Pin17,Sun19,Cha23} have drawn model-dependent conclusions on whether synchronisation enhances~\cite{Zha20,Gui24} or reduces~\cite{Izu16,Her18} dissipation. Our results now show that synchronisation \textit{reduces} dissipation in \textit{all} driven Potts models, and that the least dissipative non-equilibrium states are dynamically selected close to the phase transition.
\section{Macroscopic dynamics}
From a minimal set of requirements, including thermodynamic consistency, symmetry, and a well-defined thermodynamic limit, we show in Ref.~\cite{Mei24b}, that as $N\to\infty$ the occupation probabilities $p_n(t)$ obey the deterministic equations of motion
\algn{\eqnlab{mf}
	\dd{t} p_n(t) 	\equiv	 h_n[\ve p(t)] =  j(p_{n},p_{n-1}) - j(p_{n+1},p_n)\,,
}
$n=0,\ldots,q-1$, where $j(p_{n+1},p_n)$ denotes the average probability flux per oscillator from state $n$ to state $n+1$. The flux $j$ is expressed in terms of the rescaled microscopic transition rates $w_n^\pm$ for the transitions $s_{m}\to s_{m}\pm1$, $s_m=n$, as
\algn{\eqnlab{flux}
	j(p_{n+1},p_n) 	= w_n^+(\ve p ) - w^-_{n+1}(\ve p)\,.
}
The LDB condition~\cite{Sei12,Bro15,Pel21} requires that
\algn{
	\frac{w^\pm_n(\ve p)}{w^\mp_{n\pm1}(\ve p )}	= \exp\left\{-\beta \left[(\partial_{p_{n\pm1}} - \partial_{p_n})\mc{F}(\ve p)\mp f\right]\right\}\,,
}
where $f$ denotes the non-conservative force and
\algn{
	\mc{F}(\ve p)			=  -\frac{\ms{J}}{2}\ve p\cdot\ve p + \beta^{-1}\ve p\cdot\log \ve p\,,
}
denotes the free energy per oscillator.

From \Eqnref{mf}, we immediately find that the decoherent state $\ve p^*$ is a fixed point with $\dd{t}\ve p|_{\ve p = \ve p^*} = 0$. A linear stability analysis reveals that the stability of $\ve{p}^*$ depends on the sign of the bifurcation parameter~\cite{Mei24b}
\algn{
	\Lambda = 2(j_{10}-j_{01})\,,
}
where we denote by $j_{nm}$ the derivatives of the probability flux $j$ at the decoherent fixed point:
\algn{
	j_{nm} \equiv \partial^n_{y}\partial^m_xj(y,x)|_{x=y=\frac1q}\,.
}
The fixed point $\ve p^*$ is stable for $\Lambda<0$ and unstable for $\Lambda>0$, so that perturbations of decoherence grow, whenever $\Lambda$ is positive.

The location in parameter space where $\Lambda$ changes sign depends on the transition rates. Figure~\figref{PD}(a) shows the phase boundaries ($\Lambda=0$) for different $w^\pm_n$. In the equilibrium limit $\beta f \to 0$, all phase boundaries approach $\beta\!\!\ms{J}_c/q = 1$~\cite{Wu82}. Away from equilibrium, for $\beta f>0$, varying $w^\pm_n$ may either stabilise the decoherent phase (blue arrow), destabilise it (red arrow), or retain the phase boundary of the equilibrium case (solid line), see Ref.~\cite{Mei24b} for examples.

For $\Lambda>0$, when the decoherent fixed point is unstable, fluctuations drive the system away from decoherence and towards other attractive states. Such states can be either distant attractors such as ordered states, as is the case at equilibrium~\cite{Wu82} and for weak driving~\cite{Her18}, or nearby, small-amplitude variations of $\ve p^*$. Candidates for the latter are coherent oscillations (synchronisation) and stationary, non-equilibrium patterns. 
\begin{figure}
	\includegraphics[width=\linewidth]{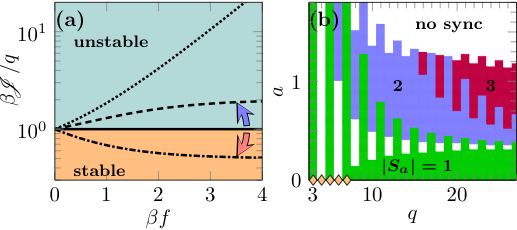}
	\caption{(a) Stability of decoherent phase in the $\beta f$-$\beta\!\!\ms{J}/q$ plane with phase boundaries for different dynamics (broken and solid lines, details in Ref.~\cite{Mei24b}). (b) Phase diagram of small-amplitude synchronised states in driven Potts models as function of $q$ and $a$ [\Eqnref{amat3}]. Numbers indicate how many Fourier modes are active. Markers show the parameters of previous numerical studies~\cite{Her18,Her19}.
	}\figlab{PD}
\end{figure}
\section{Analytic solution}
In order to explore the possible stable structures in the vicinity of the decoherent fixed point $\ve p^*$, we solve the driven Potts model exactly for small $\ve x$. To this end, we exploit the periodicity of Potts spins and transform the variation vector $\ve x$ into discrete Fourier modes $\ve{\hat x}$ by means of a Fourier transform $\ve{\hat x} = \mbb{F}\ve x$, where $\mbb{F}$ is a $q\times q$ matrix with elements $F_{kn} = \exp(i2\pi k n/q)$, $k=0,\ldots,q-1$. The inverse transform $\mbb{F}^{-1}$ is obtained by Hermitian conjugation: $\mbb{F}^{-1} = q^{-1}\mbb{F}^{\dagger}$.

This way, we obtain the equation of motion for $\ve{\hat x}$:
\algn{\eqnlab{eomz}
	\dot{\ve{\hat x}} \equiv  \ve{\hat h}(\ve{\hat x})\,, \qquad  \ve{\hat h}(\ve{\hat x})\equiv \mbb{F}\ve{h}(\mbb{F}^{-1}\ve{\hat x}+\ve p^*)\,.
}

In the vicinity of the decoherent fixed point, $|\ve{\hat x}| \ll 1$, we expand $\ve{\hat h}(\ve{\hat x})$ to third order in $\ve{\hat x}$, i.e.,
\algn{\eqnlab{eomexp}
	\ve{\hat h}(\ve{\hat x}) \sim \ve{\hat h}^{(1)}\!\!(\ve{\hat x}) + \ve{\hat h}^{(2)}\!\!(\ve{\hat x}) + \ve{\hat h}^{(3)}\!\!(\ve{\hat x})		\,,
}
where $\ve{\hat h}^{(n)}$, $n=1,\ldots,3$, are $n^\text{th}$-order polynomials in $\ve{\hat x}$.

The linear part $\ve{\hat h}^{(1)} = \mbb{D}\ve{\hat x}$ contains the stability matrix $\mbb{D}$, diagonal in the Fourier basis, with elements $\mbb{D}_{kk} = \mu_k + i\omega_k$ and
\algn{
	\mu_k = \Lambda\sin^2\left(\frac{\pi k}{q}\right)\,,\quad \omega_k =  \Omega\sin\left(\frac{2\pi k}{q}\right)\,.
}
The imaginary part $\omega_k$ of $\mbb{D}_{kk}$, with parameter
\algn{
	\Omega = j_{10}+j_{01} = qj_{00}\,,
}
denotes the ``natural frequency'' of the $k^\text{th}$ Fourier mode, with which the dynamics spirals away from $\ve p^*$ for $\Lambda>0$~\cite{Mei24b}.

Where the dynamics takes $\ve{\hat x}(t)$ in the long-time limit depends on the higher-order terms in \Eqnref{eomexp}. Before we consider these, we note that $\ve x$ is real, so that the Fourier modes $\ve{\hat x}$ are related by complex conjugation. We define
\algn{\eqnlab{zdef}
	z_k \equiv {\hat x}_k\,, \qquad \bar z_k \equiv {\hat x}_{-k}\,,
}
with $k=1,\ldots,\lfloor\frac{q}2\rfloor$ (indices modulo $q$). The zeroth mode $z_0=\hat x_0$ vanishes due to probability conservation, leaving the driven $q$-state Potts model with $\lfloor\frac{q}2\rfloor$ dynamic Fourier modes, as stated for $q=7$ in the Introduction.

The non-linear terms in \Eqnref{eomexp} are brought into the simpler normal form~\cite{Mei24b}
\algn{\eqnlab{bhnf}
	\dot{z}'_{k} \sim \bigg(\mbb{D}_{kk} - \sum_{k=1}^{\lfloor \frac{q}2\rfloor}\mbb{C}_{kk'}|z'_{k'}|^2 \bigg) z'_k\,,
}
by a non-linear transformation $z_k\mapsto z'_k$, where $\mbb{C}$ is a complex matrix. The normal form~\eqnref{bhnf} retains only terms that are essential for the transition when $0<\Lambda\ll1$ and $\Lambda\ll\beta f$ \footnote{These limits should be understood as conservative mathematical statements. We find numerically that \Eqsref{bhnf}, \eqnref{dsig}, and \eqnref{dl} hold also for moderate $\Lambda$ and $\beta f$}.

To determine the long-time states, we separate the Fourier modes $z'_k$ into their amplitudes $r_k$ and phases $\phi_k$, $z'_k = r_k \exp(i\phi_k)$. The normal form~\eqnref{bhnf} then decomposes into a phase equation for $\ve{\phi}$ and an amplitude equation for $\ve{r}$. The latter reads
\sbeqs{
\algn{\eqnlab{amp}
	\dot r_k \sim \big(\mu_k - \sum_{k'=1}^{\lfloor \frac{q}2\rfloor}\mbb{A}_{kk'}r^2_{k'}\big)r_k
}
with~\cite{Mei24b}
\algn{\eqnlab{amat1}
	\mbb{A}\equiv \text{Re}(\mbb{C}) = A\left(\ve u \ve v^{\sf T} - \mbb{W}\right)\,,\quad A \equiv \frac{4}{q^2}(j_{03}-j_{30})\,,
}
}
and contains the essential parameters for $\ve r$ close to the transition. The vectors $\ve u$, $\ve v$ and the diagonal matrix $\mbb{W}$ depend only on the single parameter
\algn{\eqnlab{amat3}
	a = \frac{j_{02}^2-j_{20}^2}{q j_{00}(j_{03}-j_{30})} - 1\,.
}
Both stable synchronised states and stationary patterns with $\ve{\dot\phi}=0$ are given by the stable fixed points $\ve{r}^*$ of \Eqnref{amp}, i.e., $\ve{\dot r}|_{\ve r = \ve{r}^*} = 0$. Because of the simple structure~\eqnref{amat1} of $\mbb{A}$, these fixed points can be obtained explicitly~\cite{Mei24b}.

Following this program, we arrive at the phase diagram of small-amplitude synchronised states in the $q$-$a$ parameter space, given in \Figref{PD}(b). In the white regions, no synchronised states are stable. Different colours indicate stable synchronisation patterns with a finite number $|S_a|$ of non-vanishing, i.e. active, Fourier modes. The amplitudes of the remaining $\lfloor \frac{q}{2} \rfloor - |S_a|$ modes vanish, i.e. these modes are inactive.

The green region ($|S_a|=1$) exhibits multistability, i.e., several synchronised states, each with a different active mode, are simultaneously stable~\cite{Mei24b}. The other coloured regions with $|S_a|>1$ host unique stable synchronised states.

The white patches between the green and blue regions in \Figref{PD}(b) occur only for even $q$. Although they host no synchronised states, they admit stationary probability patterns~\cite{Mei24b}. Note that previous numerical studies~\cite{Her18,Her19}, shown as markers in \Figref{PD}(b), covered only a tiny fraction of the parameter space.
\section{Stability-dissipation relation}
We now derive the main result of this Letter, a stability-dissipation relation, that holds for all dynamics that admit stable small-amplitude states, including both synchronisation and stationary patterns. To this end, we introduce the average entropy production rate $\langle\dot\sigma\rangle$ per oscillator~\footnote{We express all entropy measures in dimensionless form, i.e., in units of $\kb$.}, which for $N\to\infty$ takes the form~\cite{Mei24b}
\algn{\eqnlab{entprod}
	 \langle\dot\sigma\rangle  = \sum_{n=0}^{q-1} F(p_{n+1},p_n) j(p_{n+1},p_n)\,,
}
where $F(p_{n+1},p_n) = \beta\left[ f - (\partial_{p_{n+1}}-\partial_{p_n})\mc{F}(\ve p) \right]$ are thermodynamic forces that drive the fluxes $j(p_{n+1},p_n)$.

We define the change in entropy production $\Delta \dot\sigma$ relative to the entropy production rate of a single Potts oscillator $\langle\dot\sigma\rangle_0\equiv 2\beta f\sinh(\beta f/2)$~\cite{Her18,Her19} by $\Delta \dot\sigma = (\langle \dot\sigma \rangle-\langle \dot\sigma \rangle_0)/|\langle \dot\sigma \rangle_0|$, express $\langle\dot\sigma\rangle$ in terms of Fourier modes, and expand around the decoherent fixed point $\ve{\hat x}=0$.  Exploiting the symmetries of driven Potts models and the general properties of the transition rates, we obtain the simple expression~\cite{Mei24b}:
\algn{\eqnlab{dsig}
	\Delta \dot\sigma \sim -2\Gamma\lambda \alpha_a<0\,, \qquad \Gamma \equiv \frac{2j_{11}}{q^2 j_{00}}\,,
}
valid for $0<\Lambda\ll1$ and $\Lambda\ll\beta f$~\cite{Note1}. The parameters $\lambda \equiv \Lambda/A$ and $\Gamma>0$ are dynamics dependent, while $\alpha_a>0$ depends on the set of active Fourier modes in a given stable state and on $a$~\cite{Mei24b}. Equation~\eqnref{dsig} shows that entropy production is reduced in any stable small-amplitude state.

Finally, we connect $\Delta \dot\sigma$ to the stability change across the transition. To this end, we consider the average $\langle L\rangle$ of the stochastic inflow rate $L$~\cite{Bai15}, the difference between the entrance rate, the sum of all transition rates of adjacent states entering a given state, and the escape rate out of this state.

In the thermodynamic limit, $\langle L \rangle$ converges to the phase-space contraction rate $\mc{L}$, i.e., $\lim_{N\to\infty}\langle L\rangle = \mc{L}$~\cite{Mei24b}, given by $\mc{L} = -{\ve \nabla}_{\!\!\ve p}\cdot\dot{\ve p}$~\cite{Ott02}. Phase-space probability accumulates in regions of positive $\mc{L}$, while it escapes regions of negative $\mc{L}$. Positive $\mc{L}$ is a necessary condition for fixed points and periodic orbits to be approached in the long-time limit.

We analyse the relative stability change $\Delta\mc{L} =(\mc{L}-\mc{L}_0)/|\mc{L}_0|$ of $\mc{L}$ close to the transition, where $\mc{L}_0 \equiv -q\Lambda/2$ denotes the phase-space contraction rate at $\ve p^*$. Expanding $\Delta\mc{L}$ around $\ve{\hat x}=0$, we find~\cite{Mei24b}

\algn{\eqnlab{dl}
	\Delta\mc{L} \sim 2\alpha_a >0\,,
}
implying that small-amplitude states are more stable ($\mc{L}$ is larger) than $\ve p^*$ for $\Lambda>0$.

Combining \Eqnref{dl} with \Eqnref{dsig}, we arrive at the stability-dissipation relation
\algn{\eqnlab{sdr}
	\Delta\dot\sigma \sim -\Gamma \lambda \Delta\mc{L}\,,
}
the main results of this Letter. It shows that, close to the transition, $\Delta\dot\sigma$ depends linearly on $\Delta\mc{L}$ with negative, dynamics-dependent prefactor $-\Gamma \lambda$, so that dissipation is smallest in the most stable small-amplitude state.
\begin{figure}
	\includegraphics[width=\linewidth]{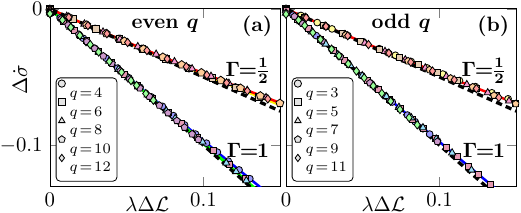}
\caption{Stability-dissipation relation for two different dynamics with $N=10^6$ and $\beta f=7$, obtained from a time average with $t/\tau \approx 100$. (a) Even $q$. Markers from numerical simulations of the stochastic dynamics, solid lines from the macroscopic dynamics~\eqnref{mf}. The dashed line shows \Eqnref{sdr}. (b) Same as in (a) but for odd $q$.}\figlab{sdr}

\end{figure}

We have tested the stability-dissipation relation numerically, using both the macroscopic~\eqnref{mf} and the stochastic dynamics.  The latter involves an additional long-time average~\cite{Mei24b}. Figure~\figref{sdr} shows $\Delta\dot\sigma$ plotted against $\lambda\Delta\mc{L}$ for two different sets of transition rates with $(a,\Gamma)=(0,1/2)$ and $(a,\Gamma)=(1,1)$~\cite{Mei24b}. For values of $\Delta\dot\sigma$ and $\lambda\Delta\mc{L}$ close to the transition, the simulations confirm \Eqnref{sdr}.
\section{Minimum-dissipation principle}
Finally, we establish the minimum-dissipation principle for driven Potts models. When the thermodynamic limit is taken before a long-time limit, \Eqnref{sdr} alone does not directly imply such a principle, because the infinite-size system is in general non-ergodic and the most stable states are assumed only if the initial condition lies in their basins of attraction. However, when long times are invoked first, fluctuations enable frequent transitions between different states that are sufficiently close. This holds for $\lambda\ll1$, where all small-amplitude states and the decoherent state are separated only by small distances of order $\sqrt{\lambda}$, while finite-size fluctuations are of order $1/\sqrt{\mc{L}N/q}\propto1/\sqrt{\lambda N}$. Hence, stable small-amplitude states are frequently visited when $1\ll N\lesssim \lambda^{-2}$.

In this case, the system predominantly occupies stable states with largest $\langle L\rangle$, because once a (stochastic) trajectory has reached such a state, it typically spends a long time there, compared to the average entrance time. Equation~\eqnref{sdr} then implies that driven Potts models probabilistically select small-amplitude states with largest $\langle L(\ve N)\rangle\to\mc{L}$ and thus smallest dissipation~\footnote{See Ref.~\cite{Mei24b} and the Supplemental Material~\cite{Supp} for numerical confirmations at $q=9$ and $q=17$.}. In numerical simulations, we find that the minimum-dissipation principle holds for substantially larger $N$ and $\lambda$ than our argument suggests~\cite{Mei24b}, which is why $1\ll N\lesssim \lambda^{-2}$ should be considered a conservative estimate.
\section{Conclusion}
We have proved a linear stability-dissipation relation~\eqnref{sdr} for driven Potts models, which shows that the most stable dissipative structures dissipate the least entropy. For large but finite systems, we argued that \Eqnref{sdr} implies a minimum-dissipation principle for driven Potts models, that holds arbitrarily far away from equilibrium, independently of the dynamics, and for all $q$. The dissipation and stability measures $\dot\sigma$ and $L$ can be defined for any thermodynamically consistent stochastic system. We are therefore confident that the tools developed here will facilitate further connections between non-equilibrium thermodynamics and stability. For instance, preliminary results suggest a relation similar to \Eqnref{sdr} for Potts oscillators with local interactions.
\section{Acknowledgments}
\begin{acknowledgments}
	We thank G. Falasco for insightful discussions and for pointing out the notion of phase-space contraction in stochastic systems.
	J.M.'s stay at King's College London was supported by a Feodor-Lynen scholarship of the Alexander von Humboldt-Foundation and M.E. by the ChemComplex project (C21/MS/16356329) funded by FNR (Luxembourg).
\end{acknowledgments}
\end{document}